\documentclass[%
 aip,
 amsmath,amssymb,
reprint,%
]{revtex4-1}

\usepackage{graphicx}
\usepackage{dcolumn}
\usepackage{bm}
\usepackage{comment}
\usepackage[utf8]{inputenc}
\usepackage[T1]{fontenc}
\usepackage{mathptmx}
\usepackage{etoolbox}
\usepackage{braket}
\usepackage{titlesec}
\usepackage{caption}
\usepackage{subcaption}
\usepackage{xspace} 

\usepackage{xr-hyper}
\usepackage{hyperref}
\usepackage{cleveref}

\makeatletter
\def\@email#1#2{%
 \endgroup
 \patchcmd{\titleblock@produce}
  {\frontmatter@RRAPformat}
  {\frontmatter@RRAPformat{\produce@RRAP{*#1\href{mailto:#2}{#2}}}\frontmatter@RRAPformat}
  {}{}
}%
\makeatother
\begin{document}

\preprint{AIP/123-QED}

\title[The Vertical Excitation Model (VEM) within Polarizable QM/MM]{Vertical Excitation Energies of Embedded Systems: The Vertical Excitation Model (VEM) within Polarizable QM/MM}
\author{Chiara Sepali}
    \affiliation{Scuola Normale Superiore, Piazza dei Cavalieri 7, 56126 Pisa, Italy.}%
\author{Piero Lafiosca}
\affiliation{Scuola Normale Superiore,
	Piazza dei Cavalieri 7, 56126 Pisa, Italy.}
\author{Linda Goletto}
\affiliation{Scuola Normale Superiore,
	Piazza dei Cavalieri 7, 56126 Pisa, Italy.}
\author{Tommaso Giovannini}
\affiliation{Department of Physics, University of Rome Tor Vergata, and INFN, Via della Ricerca Scientifica 1, 00133, Rome, Italy.}    
\author{Chiara Cappelli}
\affiliation{Scuola Normale Superiore,
	Piazza dei Cavalieri 7, 56126 Pisa, Italy.}
\email{chiara.cappelli@sns.it}

\date{\today}

\begin{abstract}
Polarizable Quantum Mechanics/Molecular Mechanics (QM/MM) approaches based on fluctuating charges and dipoles (QM/FQ(F$\mu$)) are formulated within the state-specific Vertical Excitation Model (VEM) to compute vertical excitation energies of solvated systems. This methodology overcomes the limitations of the widely used Linear Response (LR) approach. While LR can capture the dynamic response of the solvent to the QM transition density, it neglects the solvent reorganization that follows solute relaxation upon electronic excitation. In contrast, the VEM framework explicitly accounts for this effect. Benchmark calculations of vertical excitation energies using QM/FQ(F$\mu$) are reported for a representative set of solutes—acrolein, acetone, caffeine, p-nitroaniline, coumarin 153, doxorubicin, and betaine-30—comparing VEM with LR, corrected LR (cLR), and cLR$^2$ schemes. The results reveal notable variations in solvent response depending on the character of the electronic transition and demonstrate that optimal accuracy can be achieved by selecting the most appropriate model for each specific system and excitation.
\end{abstract}

\maketitle

\section{Introduction}

Computational chemistry has long relied on embedding models to study molecular systems interacting with an external environment. \cite{tomasi1994molecular,tomasi2005quantum,warshel1976theoretical,warshel1972calculation,orozco2000theoretical,giovannini2020molecular,knizia2013density,wesolowski1993frozen}
Among these, quantum mechanics/molecular mechanics (QM/MM) methods have rapidly become the state-of-the-art for describing strongly interacting complex systems, owing to their ability to account for interactions between a system and its surroundings at an atomistic level. \cite{warshel1972calculation,warshel1976theoretical,senn2009qm,lin2007qm,giovannini2020molecular,gomez2022multiple,bondanza2020polarizable}
For solvated systems, the most widely employed QM/MM strategy treats the solute at the QM level while representing the solvent molecules through classical MM force fields. The quality of QM/MM results for a given QM level is thus strongly dependent on the accuracy of the methodology used to describe the interactions between the two subsystems. Typically, only electrostatic contributions are explicitly considered, while non-electrostatic effects are represented through empirical functions, such as Lennard-Jones potentials, and are therefore neglected in the evaluation of response properties. \cite{giovannini2020molecular,gomez2022multiple,loco2016qm,bondanza2020polarizable,giovannini2017general}
Focusing on electrostatic interactions, the most common representation of the MM region assigns fixed atomic charges, defining the so-called Electrostatic Embedding (EE) model. \cite{senn2009qm} In this scheme, the MM layer polarizes the electronic density of the QM region, but the reciprocal effect is absent. However, a more physically accurate description of solute–solvent interactions requires accounting for mutual polarization. To this end, several polarizable embedding frameworks have been proposed, which differ in how electrostatic and polarization effects are treated, thereby influencing both the solute electronic structure and its response to external perturbations. \cite{loco2016qm,thole1981molecular,curutchet2009electronic,olsen2011molecular,steindal2011excitation,cappelli2016integrated,giovannini2019polarizable,boulanger2012solvent,rick1994dynamical,thompson1995excited,dziedzic2016tinktep}
In polarizable QM/MM models, the MM polarization quantities—typically charges and/or dipoles—respond self-consistently to the QM density and vice versa, establishing a two-way interaction. This mutual coupling significantly improves the accuracy in describing complex embedded systems and is particularly effective for the computation of spectroscopic properties, especially when combined with extensive configurational sampling. \cite{loco2016qm,curutchet2009electronic,olsen2011molecular,steindal2011excitation,cappelli2016integrated,giovannini2019polarizable,boulanger2012solvent,thompson1995excited} 

Polarizable QM/MM approaches are of paramount importance for accurately modeling electronic absorption spectra in complex environments such as solutions or biological systems. \cite{giovannini2020molecular,gomez2022multiple,bondanza2020polarizable,loco2016qm}
By treating the chromophore at the quantum mechanical level and representing the surrounding environment with a classical yet polarizable model, these methods provide an optimal balance between accuracy and computational efficiency. \cite{giovannini2020molecular,gomez2022multiple,bondanza2020polarizable,loco2016qm}
They enable the explicit inclusion of polarization effects, specific solute–solvent interactions, and dynamic fluctuations, all of which strongly influence spectral features. In particular, accounting for mutual polarization is crucial to quantitatively reproduce solvatochromic shifts and spectral line shapes. \cite{giovannini2020molecular,gomez2022multiple,bondanza2020polarizable,loco2016qm}
Among the different polarizable embedding schemes, models based on fluctuating charges and dipoles (QM/FQ(F$\mu$))\cite{cappelli2016integrated,giovannini2019polarizable}have emerged as a particularly effective strategy, as they allow a consistent treatment of mutual solute–solvent polarization within both ground- and excited-state calculations.

An important aspect when using embedding approaches to model absorption and emission spectra is achieving a physically consistent description of how the environment affects excitation and de-excitation processes.
In the gas phase, excitation and emission energies can be computed using either Linear Response (LR) \cite{casida1995time} or State-Specific (SS) approaches, which are equivalent in the limit of the exact solution of the corresponding equations. \cite{caricato2006formation,cammi2005electronic,corni2005electronic}
In polarizable QM/MM approaches, however, the presence of an additional non-linear term in the QM Hamiltonian (the so-called reaction field) introduces a crucial distinction between the two methods, as they describe fundamentally different solvent responses following electronic excitation. \cite{cammi2005electronic,corni2005electronic,caricato2006formation}
Specifically, LR models the response of the ground-state (GS) solution to time-dependent electric fields, and the absorption spectrum is obtained in a single step by evaluating the poles of the appropriate response function. \cite{casida1995time}
LR is the most widely employed method for simulating vertical excitation energies in the condensed phase, as it provides an excellent balance between computational cost and accuracy. \cite{corni2005electronic,cammi2005electronic,cupellini2015electronic,guido2015electronic,gomez2022multiple,bondanza2020polarizable,cappelli2016integrated}
Within this framework, the solvent is polarized by the QM transition density, producing an in-phase reaction field that acts back on the transition density. This dynamic solvent polarization can be interpreted as a dispersion-like interaction. \cite{corni2005electronic,cammi2005electronic,guido2015electronic}
Numerous polarizable QM/MM approaches have been extended to the LR formalism within Time-Dependent Density Functional Theory (TDDFT). \cite{loco2016qm,curutchet2009electronic,lipparini2012linear,steindal2011excitation,rinkevicius2014hybrid,jensen2003discrete,cappelli2016integrated,giovannini2019electronic,giovannini2019calculation}
This formulation has proven highly effective in capturing environmental effects for excitations involving bright states with large transition dipole moments.
However, it fails to describe the solvent relaxation that accompanies substantial changes in the QM density during excitation, such as those occurring in charge-transfer (CT) transitions. \cite{guido2015electronic,loco2016qm,giovannini2019electronic}
 
A natural way to overcome the intrinsic limitations of LR is to adopt an SS framework; various SS methods within the TDDFT formalism have been developed for different embedding approaches. \cite{caricato2006formation, improta2006state, marenich2011practical,caricato2014corrected,guido2017excited,schroder2018corrected,loco2016qm,giovannini2019electronic}
Among these, the corrected LR model (cLR), originally formulated for the implicit Polarizable Continuum Model (PCM) and later extended to fully atomistic QM/MM methods, \cite{caricato2006formation,giovannini2019electronic,loco2016qm} introduces a first-order perturbative SS correction into the LR scheme. Specifically, the cLR model partially accounts for the SS response of the solvent to a given excited-state (ES) solute density. While computationally efficient, cLR captures only a fraction of the SS effects when the excitation induces substantial changes in the QM electron density. An alternative is the so-called cLR$^2$ scheme, \cite{guido2021simple} which combines LR and first-order SS corrections at the same computational cost as cLR, thus including both dynamic (dispersion, LR) and static (SS, cLR) components of the solvent response.
More recently, the implicit PCM has been formulated within the self-consistent Vertical Excitation Model (VEM), providing a complete and robust framework for incorporating SS effects. \cite{marenich2011practical,guido2017excited,bjorgaard2015solvent}
In VEM, the solvent response is driven by the density difference between the GS and the ES of interest, either unrelaxed (\textbf{T}) or relaxed ($\textbf{P}^\Delta$), thus leading to the VEM(\textbf{T}) and VEM($\textbf{P}^\Delta$) formulations. \cite{marenich2011practical}
The adoption of an SS solvation scheme is essential for accurately capturing solvent reorganization following solute relaxation after electronic excitation. As a result, SS models are expected to provide a more physically faithful description of the process compared to LR. \cite{corni2005electronic,lunkenheimer2013solvent,marenich2011practical,guido2015electronic,jacquemin2011excited}
Another key advantage of VEM is the availability of analytic gradient formulations based on a Lagrangian approach, \cite{bjorgaard2015solvent,guido2017excited} which enable excited-state molecular dynamics and geometry optimization, that remain challenging for other SS models such as cLR, where gradient calculations are cumbersome and computationally demanding.

In this work, the Vertical Excitation Model (VEM) is introduced for the first time within polarizable QM/MM approaches and, in particular, extended to the QM/Fluctuating Charges (FQ) \cite{rick1994dynamical,cappelli2016integrated} and QM/Fluctuating Charges and Fluctuating Dipoles (FQF$\mu$) \cite{giovannini2019polarizable} models, which have previously been formulated within the LR, cLR, and cLR$^2$ frameworks. \cite{giovannini2019electronic,ambrosetti2021quantum,lipparini2012linear}
The performance of LR, cLR, cLR$^2$, and VEM is assessed by calculating QM/FQ(F$\mu$) vertical excitation energies for molecules exhibiting different types of electronic transitions. Solvation dynamics and accurate sampling of the solute–solvent configurational space are ensured by coupling QM/FQ(F$\mu$) with molecular dynamics (MD) simulations, following protocols established in previous studies. \cite{giovannini2020molecular,gomez2022multiple}

The paper is organized as follows. The next section provides a brief overview of the QM/FQ(F$\mu$) models and highlights the main features of the LR, cLR, cLR$^2$, and VEM approaches, with VEM then formulated explicitly for QM/FQ(F$\mu$). This is followed by a concise description of the computational protocol and calculation details. The various SS approaches are subsequently applied, and the results are compared with both gas-phase calculations and available experimental data. The paper concludes with a discussion of the main findings and perspectives for future work.

\section{Theory}

\subsection{QM/FQ and QM/FQF\texorpdfstring{$\mu$}{mu} polarizable embedding approaches}

In the FQ \cite{rick1994dynamical,rick1995fluctuating,cappelli2016integrated} and FQF$\mu$ \cite{giovannini2019polarizable} force fields, each MM atom is assigned either a charge $q$ (in the case of FQ) or both a charge $q$ and a dipole $\pmb{\mu}$ (in the case of FQF$\mu$), which adjust according to the external potential. For a system composed of various molecules, the FQF$\mu$ energy functional is expressed as follows:\cite{cappelli2016integrated,giovannini2019polarizable}
\begin{equation}
\begin{aligned}\label{eq:funct}
E_{\text{FQ(F$\mu$)}} & = \sum_{i \alpha} \text{q}_{i \alpha} \chi_{i \alpha} + \frac{1}{2} \sum_{i \alpha} \sum_{j \beta} \text{q}_{i \alpha} \text{T}_{i \alpha, j \beta}^{\text{qq}} \text{q}_{j \beta} \\ 
& \quad + \sum_{i \alpha} \sum_{j \beta } \text{q}_{i \alpha} \textbf{T}_{i \alpha,j \beta}^{\text{q} \mu} \pmb{\mu}_{j \beta} + \frac{1}{2} \sum_{i \alpha} \sum_{j \beta } \pmb{\mu}_{i \alpha}^\dagger \textbf{T}_{i \alpha, j \beta}^{\mu \mu} \pmb{\mu}_{j \beta} \\
& \quad + \sum_{\alpha} \Big{[} \lambda_\alpha \sum_i (\text{q}_{i \alpha}) - \text{Q}_{\alpha} \Big{]}
\end{aligned}
\end{equation}
where the first two terms represent the FQ energy, which is recovered by removing the contributions associated with the dipoles $\pmb{\mu}$. ($i, j$) and ($\alpha, \beta$) indices run over FQ atoms and molecules, respectively. $\text{T}_{i \alpha, j \beta}^{\text{qq}}$, $\textbf{T}_{i \alpha, j \beta}^{\text{q}\mu}$, and $\textbf{T}_{i \alpha, j \beta}^{\mu \mu}$ are the charge--charge, charge--dipole, and dipole--dipole interaction kernels. \cite{cappelli2016integrated, giovannini2019polarizable} 
To avoid the so-called ``polarization catastrophe'', \cite{thole1981molecular} the FQ force field employs the Ohno kernel \cite{cappelli2016integrated,ohno1964some}, whereas the FQF$\mu$ model adopts the Gaussian kernel.\cite{giovannini2019polarizable} $\chi_{i \alpha}$ is the atomic electronegativity. 
The diagonal elements $T_{i\alpha,i\alpha}^{qq}$ are defined according to the atomic chemical hardness $\eta_{i\alpha}$, while in the FQF$\mu$ model, $T_{i\alpha,i\alpha}^{\mu \mu}$ are associated with the atomic polarizability $\alpha_{i \alpha}$. Consequently, the atomic parameters for the FQ and FQF$\mu$ models include the electronegativity $\chi_{i \alpha}$, the chemical hardness $\eta_{i \alpha}$, and, specifically for the FQF$\mu$, the atomic polarizability $\alpha_{i \alpha}$. 
To prevent unphysical charge transfer between FQ(F$\mu$) molecules, \cite{chen2009dissociation} a set of Lagrangian multipliers $\lambda_\alpha$ constrains each molecule to retain a total charge $\text{Q}_\alpha$. 

The equilibrium charges--and dipoles for FQF$\mu$--are obtained by satisfying the Electronegativity Equalization Principle (EEP), \cite{sanderson1951interpretation} which corresponds to minimizing the energy functional in \cref{eq:funct} with respect to charges, dipoles, and Lagrangian multipliers, leading to the following linear system: \cite{cappelli2016integrated,giovannini2020molecular}
\begin{equation}
\left(
\begin{array}{cc|c}
\mathbf{T}^{\text{qq}} & \mathbf{1}_{\lambda}  & \mathbf{T}^{\text{q}\mu} \\ 
\mathbf{1}^{\dagger}_{\lambda} & \mathbf{0} & \mathbf{0}  \\
\hline
\mathbf{T}^{\text{q}\mu^{\dagger}} & \mathbf{0} & \mathbf{T}^{\mu\mu}
\end{array}
\right)
\left({\begin{array}{c} 
\mathbf{q}\\
\bm{\lambda}\\ 
\hline
\bm{\mu} 
\end{array}}\right)
=
\left(\begin{array}{c} -\bm{\chi} \\ 
\mathbf{Q}_{\alpha} \\
\hline
\mathbf{0}
\end{array}\right) 
\label{eq:MMlinearsys1}
\end{equation}
where $\mathbf{1}_{\lambda}$ denotes rectangular blocks containing the Lagrangian multipliers, and the overbars are used to separate FQ contributions from the additional terms of FQF$\mu$.

FQ(F$\mu$) can be coupled to a QM description of a subsystem (e.g. the solute in a solution) in a QM/MM framework. The total energy can therefore be written as the sum of three different terms: \cite{senn2009qm, giovannini2020molecular, cappelli2016integrated, giovannini2019polarizable}
\begin{equation}\label{eq:totalenergy}
    E = E_{\text{QM}} + E_{\text{FQ(F$\mu$)}} + E^{\text{int}}_{\text{QM/FQ(F$\mu$)}}
\end{equation}
where $E_\text{QM}$ depends on the selected level of theory, $E_\text{FQ(F$\mu$)}$ comes from \cref{eq:funct}, and $E^\text{int}_\text{QM/FQ(F$\mu$)}$ is the interaction term. 

 The QM/FQ(F$\mu$) interaction energy is given by: \cite{cappelli2016integrated, giovannini2019polarizable}
\begin{equation}
\begin{aligned}
    E^{\text{int}}_{\text{QM/FQ(F$\mu$)}} &= \sum_i q_i V [\rho_{\text{QM}}] (\mathbf{r}_i) - \sum_i {\boldsymbol{\mu}}_i \mathbf{E}[\rho_{\text{QM}}] (\mathbf{r}_i)
\end{aligned}
\end{equation}
where $V[\rho_{QM}](\textbf{r}_i)$ and $\mathbf{E}[\rho_{QM}](\textbf{r}_i)$ are the electric potential and field, respectively, generated by the QM density $\rho_{QM}$ calculated on the MM atom at position $\textbf{r}_i$. To account for the influence of the FQ(F$\mu$) layer on the electronic density, the Hamiltonian of the system is modified by the operator $\hat{H}_{\text{QM/FQ(F$\mu$)}}$, which can be written as: 
\begin{equation}\label{eq:ham}
\hat{H}_{\text{QM/FQ(F$\mu$)}} = \sum_i \frac{q_i}{|\mathbf{r}_i - \mathbf{r}|} - \sum_i \boldsymbol{\mu}_{i} \frac{\mathbf{r}_i - \mathbf{r}}{|\mathbf{r}_i - \mathbf{r}|^3}
\end{equation}
where the first term represents the potential generated at position $\mathbf{r}$ by the induced charges $q_i$ associated with each MM atom. Analogously, the second term describes the electric field generated by the fluctuating dipoles $\boldsymbol{\mu}_i$ at position $\mathbf{r}$. 

Considering the entire expression given by \cref{eq:totalenergy}, the multipoles can be obtained by imposing the global functional to be stationary with respect to charges, Lagrangian multipliers, and dipoles. The resulting linear system can be expressed as follows: \cite{cappelli2016integrated,giovannini2019polarizable}
\begin{equation}
\left(
\begin{array}{cc|c}
\mathbf{T}^{\text{qq}} & \mathbf{1}_{\lambda}  & \mathbf{T}^{\text{q}\mu} \\ 
\mathbf{1}^{\dagger}_{\lambda} & \mathbf{0} & \mathbf{0}  \\
\hline
\mathbf{T}^{\text{q}\mu^{\dagger}} & \mathbf{0} & \mathbf{T}^{\mu\mu}
\end{array}
\right)
\left({\begin{array}{c} 
\mathbf{q}\\
\bm{\lambda}\\ 
\hline
\bm{\mu} 
\end{array}}\right)
=
\left(\begin{array}{c} -\bm{\chi} \\ 
\mathbf{Q}_{\alpha} \\
\hline
\mathbf{0}
\end{array}\right) + \left(\begin{array}{c} -\mathbf{V} (\rho_{\text{QM}})\\ 
\mathbf{0} \\
\hline
\mathbf{E} (\rho_{\text{QM}})
\end{array}\right)
\label{eq:MMlinearsys2}
\end{equation}
which differs from \cref{eq:MMlinearsys1} by the inclusion of the potential $\mathbf{V} (\rho_{\text{QM}})$ and the field $\mathbf{E} (\rho_{\text{QM}})$ generated by the QM system at the MM positions on the right-hand side. Since the electronic QM density is affected by the presence of the surrounding environment (see \cref{eq:ham}) and the fluctuating charges and dipoles are in turn affected by the QM density (see \cref{eq:MMlinearsys2}), a mutually polarized system is realized.

\subsection{QM/FQ and QM/FQF\texorpdfstring{$\mu$}{mu} in the LR, cLR and cLR\texorpdfstring{$^2$}{2} regimes}

In the LR-TDDFT,\cite{lipparini2012linear, cappelli2016integrated, giovannini2019electronic,giovannini2019calculation} QM/FQ and QM/FQF$\mu$  vertical excitation energies $\omega_K$ and densities $\mathbf{X}_K, \mathbf{Y}_K$ are obtained by solving the modified Casida equations: 
\cite{lipparini2012linear, giovannini2019electronic, casida1995time, casida1998molecular}
\begin{equation}
\label{Casida}
\begin{pmatrix}
\mathbf{A} & \mathbf{B} \\
\mathbf{B^*} & \mathbf{A^*}
\end{pmatrix}
\begin{pmatrix}
\mathbf{X}_K \\
\mathbf{Y}_K
\end{pmatrix}
= \omega_K
\begin{pmatrix}
1 & 0 \\
0 & -1
\end{pmatrix}
\begin{pmatrix}
\mathbf{X}_K \\
\mathbf{Y}_K
\end{pmatrix}
\end{equation}

where A and B are modified by explicit solvent contributions, $C_{\text{ia,jb}}$, formulated as follows: \cite{lipparini2012linear, cappelli2016integrated, giovannini2019electronic}
\begin{equation}\label{eq:A}
A_{ai,bj} = (\epsilon_a - \epsilon_i)\delta_{ab}\delta_{ij} + (ai|v_{j}^{(1)} + v_{xc}^{(1)}|bj) + C^{pol}_{ai,bj}
\end{equation}
\begin{equation}\label{eq:B}
B_{ai,bj} = (ai|v_{j}^{(1)} + v_{xc}^{(1)}|bj) + C^{pol}_{ai,bj}
\end{equation}
$(i,j, \dots)$ denote occupied and $(a,b, \dots)$ virtual molecular orbitals (MOs), $v_{j}^{(1)}$ and $v_{xc}^{(1)}$ are the Coulomb and exchange-correlation potentials, respectively. $\epsilon$ indicates MO energies, while the last term in both equations, $C_{ai,bj}$, reads: \cite{giovannini2019electronic}
\begin{equation}\label{eq:casida_fqfmu}
\begin{aligned}
    C_{ai,bj} & = \sum_{i} \Big{(} \int_{\mathbb{R}^3} \phi_a (\mathbf{r}) \frac{1}{|\mathbf{r} - \mathbf{r}_i|} \phi_{i} (\mathbf{r}) d \mathbf{r} \Big{)} \cdot \text{q}_i^T (\phi_{b}, \phi_j)  \\
    & - \sum_{i} \Big{(} \int_{\mathbb{R}^3} \phi_a (\mathbf{r}) \frac{(\mathbf{r} - \mathbf{r}_i)}{|\mathbf{r} - \mathbf{r}_i|^3} \phi_{i} (\mathbf{r}) d \mathbf{r} \Big{)} \cdot \pmb{\mu}_i^T (\phi_{b}, \phi_j)
    \end{aligned}
\end{equation}
where $\mathbf{q}^{\text{T}}$ and  $\pmb{\mu}^{\text{T}}$ are the perturbed fluctuating charges and dipoles derived from the transition density $\textbf{P}_K^{\text{T}} = \mathbf{X}_K + \mathbf{Y}_K$.\cite{giovannini2019electronic} Perturbed charges $\mathbf{q}^{\text{T}}$ and perturbed dipoles $\pmb{\mu}^{\text{T}}$ are calculated, for each couple of transition vectors $\mathbf{X}_K, \mathbf{Y}_K$, by solving the following system of equations: \cite{giovannini2019electronic}
\begin{equation}
\left(
\begin{array}{cc|c}
\mathbf{T}^{\text{qq}} & \mathbf{1}_{\lambda}  & \mathbf{T}^{\text{q}\mu} \\ 
\mathbf{1}^{\dagger}_{\lambda} & \mathbf{0} & \mathbf{0}  \\
\hline
\mathbf{T}^{\text{q}\mu^{\dagger}} & \mathbf{0} & \mathbf{T}^{\mu\mu}
\end{array}
\right)
\left({\begin{array}{c} 
\mathbf{q}^T\\
\bm{\lambda}\\ 
\hline
\bm{\mu}^T 
\end{array}}\right)
=
\left(\begin{array}{c} -\mathbf{V} (\mathbf{P}_K^T)\\ 
\mathbf{0} \\
\hline
\mathbf{E} (\mathbf{P}_K^T)
\end{array}\right)
\label{eq:transition}
\end{equation}
where $\mathbf{V}(\mathbf{P_K^T})$ and $\mathbf{E}(\mathbf{P_K^T})$ are the electric potential and field due to the transition density $\mathbf{P}_K^T$. In this scheme, the solvent responds to the transition density, producing a dynamic reaction field that acts back on the system.

In the cLR approach \cite{giovannini2019electronic}, the ES relaxed density matrix of a specific state is calculated, and the excitation energy is refined to account for interactions with induced charges and dipoles, derived from the relaxed density matrix. 
In practice, cLR requires two different TDDFT calculations. \cite{mennucci2009structures, caricato2006formation} 
In the first cycle, the explicit QM/FQ(F$\mu$) contributions to the Casida equation (see \cref{eq:A,eq:B}) are set to zero. Therefore, the ES reaction field is approximated as the GS reaction field (GSRF), and the solvent affects only the GS density/orbitals. The resulting energy is denoted as $\omega_K^0$ or $\omega_{\text{GSRF}}$. The second TDDFT cycle consists of the standard QM/FQ(F$\mu$) LR-TDDFT calculation, performed by including the explicit FQ(F$\mu$) terms of \cref{eq:A,eq:B}. 
Then, the relaxed density matrix $\textbf{P}_K^\Delta$ is computed using the Z-vector approach, i.e. \cite{handy1984evaluation,giovannini2019electronic}:
\begin{equation}\label{eq:Zvector}
    \textbf{P}_K^\Delta = \textbf{P}_K^T + \mathbf{Z}_K
\end{equation}
with the Z-vector term $\mathbf{Z}_K$ accounting for orbital relaxation.
The vertical excitation energy from the GS to the $K$-th excited state reads: \cite{giovannini2019electronic}
\begin{align}
    \omega_{\text{K,FQ(F}\mu)}^{\text{cLR}} = \omega_K^0 & + \frac{1}{2} \sum_{i} q_i (\textbf{P}_K^\Delta)  V(\mathbf{r}_i, \textbf{P}_K^\Delta) \nonumber \\
    & - \frac{1}{2} \sum_{i} \pmb{\mu}_i (\textbf{P}_K^\Delta)  \mathbf{E}(\mathbf{r}_i, \textbf{P}_K^\Delta)
    \label{eq:omegaclr}
\end{align} 
Charges $q(\mathbf{P}_K^T)$ and, in the case of QM/FQF$\mu$, dipoles $\boldsymbol{\mu}(\mathbf{P}_K^T)$ are obtained through \cref{eq:transition} where the potential and field are computed by replacing $\textbf{P}_K^T$ with $\textbf{P}_K^\Delta$. 
To capture both LR and SS effects, the cLR$^2$ computational protocol described in Ref. \citenum{guido2021simple} can be applied. The cLR$^2$ vertical excitation energy is given by: \cite{guido2021simple}
\begin{align}
    \omega_{\text{K,FQ(F}\mu)}^{\text{cLR}} = \omega_K^{\text{LR}} & + \frac{1}{2} \sum_{i} q_i (\textbf{P}_K^\Delta)  V(\mathbf{r}_i, \textbf{P}_K^\Delta) \nonumber \\
    & - \frac{1}{2} \sum_{i} \pmb{\mu}_i (\textbf{P}_K^\Delta)  \mathbf{E}(\mathbf{r}_i, \textbf{P}_K^\Delta)
    \label{eq:omegaclr2}
\end{align}
This expression differs from \cref{eq:omegaclr} in the first term, where $\omega_{\text{LR}}$ replaces $\omega_{0}$. Consequently, the total excitation energy obtained with the cLR$^2$ method includes both the LR and cLR (SS) contributions, which are combined additively.\cite{guido2021simple}

\subsection{Vertical Excitation Model formulation of QM/FQ and QM/FQF\texorpdfstring{$\mu$}{mu}}

This section formulates the self-consistent, SS VEM for QM/MM models, and specifies it for QM/FQ and QM/FQF\texorpdfstring{$\mu$}{mu} force fields. Unlike the approaches discussed in the previous section, the solvent reaction field in VEM adapts to the change in the QM electronic density upon excitation. The procedure is iterative. In the initial cycle, a TDDFT calculation is performed without explicit FQ(F$\mu$) contributions in \cref{eq:A,eq:B}, thereby using the GSRF approximation to obtain the $\omega_{\text{0}}$, or equivalently $\omega_{\text{GSRF}}$, excitation energy (as the first cycle of cLR). Then the relaxed density matrix of the $K$-th ES,  $\mathbf{P}_K^\Delta$, is calculated using the aforementioned Z-vector approach, \cite{handy1984evaluation} as shown in \cref{eq:Zvector}. At the end of the first cycle, the so-called corrected GSRF (cGSRF) vertical excitation energy to the $K$-th excited state is obtained: 
\begin{align}
    \omega_{\text{K,FQ(F}\mu)}^{\text{cGSRF}} = \omega_K^0 & + \frac{1}{2} \sum_{i} q_i (\textbf{P}_K^\Delta)  V(\mathbf{r}_i, \textbf{P}_K^\Delta) \nonumber \\ 
    & - \frac{1}{2} \sum_{i} \pmb{\mu}_i (\textbf{P}_K^\Delta)  \mathbf{E}(\mathbf{r}_i, \textbf{P}_K^\Delta)
    \label{eq:cgsrf}
\end{align}
Charges $\mathbf{q}(\mathbf{P}_K^\Delta)$ and  dipoles $\pmb{\mu}(\mathbf{P}_K^\Delta)$ are computed according to \cref{eq:transition}, where the potential and field are evaluated by replacing $\textbf{P}_K^T$ with $\textbf{P}_K^\Delta$.
\cref{eq:cgsrf} is similar to \cref{eq:omegaclr}. Still, the two approaches differ: cLR includes a second TDDFT cycle to account for solvent LR terms affecting the relaxed density matrix, whereas the first iteration of VEM omits these contributions.

In the subsequent iterations of the VEM procedure, a TDDFT matrix with SS contributions is used. The TDDFT matrices $\mathbf{A}$ and $\mathbf{B}$ at the $k$-th iteration are defined as:
\begin{align}
A_{ia,jb} & = (\epsilon_a - \epsilon_i)\delta_{ab}\delta_{ij} + (ai|v_{j}^{(1)} + v_{xc}^{(1)}|bj) + \delta_{ij} \bra{a} \Delta \Phi^{(k)} \ket{b} \nonumber \\ 
& \quad - \delta_{ab} \bra{i} \Delta \Phi^{(k)} \ket{j}
\label{eq:A_VEM}
\end{align}
\begin{equation}\label{eq:B_VEM}
B_{ia,jb} = (ai|v_{j}^{(1)} + v_{xc}^{(1)}|jb) + \delta_{ij} \bra{a} \Delta \Phi^{(k)} \ket{b} - \delta_{ab} \bra{i} \Delta \Phi^{(k)} \ket{j}
\end{equation}
where occupied (i, j,...) and virtual (a, b,...) MOs are represented in the standard notation. MO energies are denoted by $\epsilon$. $\Delta \Phi^{(k)}$ is given by:
\begin{equation}
\begin{aligned}
    \Delta \Phi(\mathbf{r})^{(k)} = \sum_i \frac{q_i(\mathbf{P}_K^\Delta)}{|\mathbf{r} - \mathbf{r}_i|} - \sum_i \frac{\boldsymbol{\mu}_i(\mathbf{P}_K^\Delta)(\mathbf{r} - \mathbf{r}_i)}{|\mathbf{r} - \mathbf{r}_i|^3}  
\end{aligned}
\end{equation}
The solvent contributions that enter the TDDFT equations depend on the charges $\pmb{q}$$(\mathbf{P}_K^\Delta)$ and dipoles $\pmb{\mu}$$(\mathbf{P}_K^\Delta)$, which are derived from the relaxed density $(\mathbf{P}_K^\Delta)$ rather than the transition density $(\textbf{P}_K^{\text{T}})$, thereby including only the SS effect.
The VEM excitation energy to the $K$-th excited state at the $k$-th iteration (k>1) is expressed as:
\begin{align}
   \omega_{\text{K,FQ(F}\mu)}^{\text{VEM (k)}} = \omega_{\text{K,FQ(F}\mu)}^{\text{VEM (*k)}} & - \frac{1}{2} \sum_i q_i (\mathbf{r}_i, \textbf{P}_K^\Delta)  V(\mathbf{r}_i, \textbf{P}_K^\Delta) \nonumber \\ 
   & + \frac{1}{2} \sum_{i} \pmb{\mu}_i (\textbf{P}_K^\Delta)  \mathbf{E}(\mathbf{r}_i, \textbf{P}_K^\Delta)
\end{align}
where the first term $\omega_{\text{K,FQ(F}\mu)}^{\text{VEM (*k)}}$ is the eigenvalue of the TDDFT matrix calculated using \cref{eq:A_VEM,eq:B_VEM}. The procedure continues until convergence of the vertical excitation energy is achieved.

The formulation presented here is referred to as VEM(f,$\textbf{P}^\Delta$), although several variants of VEM exist. \cite{marenich2011practical}
For instance, if only the diagonal solvent terms of the $\mathbf{A}$ and $\mathbf{B}$ matrices in \cref{eq:A_VEM,eq:B_VEM} are retained—instead of including the full SS solvent contributions—the resulting scheme corresponds to VEM(d,$\textbf{P}^\Delta$). The rationale for considering only diagonal solvent contributions is to avoid unphysical couplings introduced by the “full” SS reaction-field operator, which can induce artificial mixing between ground and excited states through additional couplings in the occupied–occupied, virtual–virtual, and occupied–virtual subspaces. \cite{marenich2011practical} Alternatively, calculating the SS terms in \cref{eq:A_VEM,eq:B_VEM} with the unrelaxed density matrix \textbf{T} leads to the VEM(d, \textbf{T}) formulation. The unrelaxed density matrix \textbf{T}, based solely on single-excitation amplitudes, does not include orbital relaxation effects, in contrast to the relaxed one. For this reason, although computationally more demanding, the relaxed density matrix is generally preferred for computing excited-state properties. In the present work, the implemented and employed version corresponds to VEM(d,$\textbf{P}^\Delta$).

\section{Computational details}

To showcase the performance of QM/FQ and QM/FQF$\mu$ under different regimes (GSRF, LR, cLR, cLR$^2$, and VEM), a multi-step protocol is exploited. The protocol is adapted from our previous studies and aims to model the electronic properties of solutions\cite{giovannini2020molecular,gomez2022multiple}.
It involves the following steps:
\begin{enumerate}
    \item \textit{Definition of the system:} We study acrolein (ACRO), acetone (ACE), caffeine (CAFF), p-nitroaniline (PNA), coumarin 153 (C153), doxorubicin (DOXO), and betaine--30 (BET) (see \cref{fig:structures}) in aqueous solution. These molecules define the QM layer, whereas the aqueous solvent is described a the MM level, employing the FQ(F$\mu$) force fields.
  \begin{figure}
    \centering
    \includegraphics[width=0.5\textwidth]{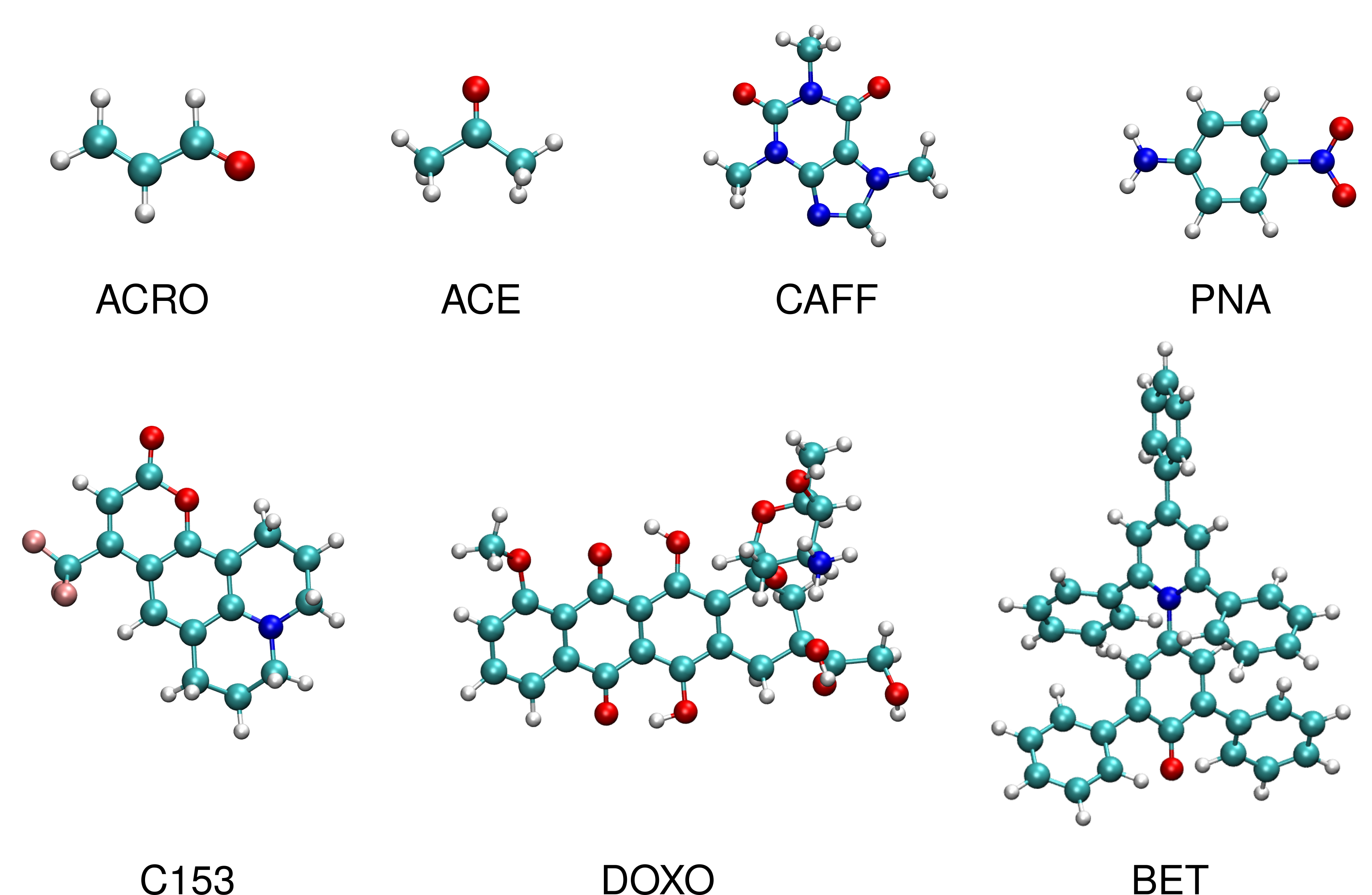}
    \caption{Molecular structures of the target molecules in aqueous solution: acrolein (ACRO), acetone (ACE), caffeine (CAFF), p-nitroaniline (PNA), coumarin 153 (C153), doxorubicin (DOXO), and betaine-30 (BET).}
    \label{fig:structures}
\end{figure}

    \item \textit{Conformational and configurational sampling:} The solute--solvent phase--space of each system is explored by performing classical MD simulations in the nanosecond time scale. 
    MD simulations of ACRO, BET, CAFF, DOXO, and ACE in aqueous solution are taken from previous studies of some of the present authors~\cite{giovannini2019quantum,ambrosetti2021quantum,gomez2020absorption,gomez2023uv,skoko2023towards}.
    A 10 ns MD simulation of aqueous C153 (NVT) using the GROMACS package \cite{abraham2015gromacs} is performed using the general AMBER force field (GAFF) \cite{wang2004development} to treat intramolecular C153 and C153-water intermolecular interactions, while TIP3P \cite{mark2001structure} is exploited for modeling water molecules (see Sec. S1 in the Supplementary Material (SM) for further details).
   \item \textit{Extraction of structures}: From the production phase of each MD simulation, 200 uncorrelated snapshots are extracted in the form of spherical droplets. Solute-centered spheres are defined with radii between 15 and 25 \AA{}, adjusted according to the solute molecule's size.  \cite{giovannini2019quantum,ambrosetti2021quantum,gomez2020absorption,gomez2023uv,skoko2023towards} Spheres with radius of 20 \AA{} are used for C153. 
    \item \textit{QM/FQ and QM/FQF$\mu$ calculations:} QM/FQ(F$\mu$) vertical excitation energies are computed for each spherical snapshot using TDDFT at the CAMY-B3LYP/TZP level to model the QM layer.\cite{te2001chemistry,nicoli2022assessing} 
    The MM layer is described with the FQ or FQF$\mu$ force fields, exploring different solvent regimes, namely the GSRF, LR, cLR, cLR$^2$, and VEM. Two different FQ parametrizations are used (FQ$^a$: see Ref.\citenum{rick1994dynamical}, FQ$^b$: see Ref. \citenum{giovannini2019effective}), while FQF$\mu$ parameters are taken from Ref.~\citenum{giovannini2019fqfmu}. For VEM calculations, VEM(d,\textbf{P}$^\Delta$) is implemented and used. All QM/MM calculations are performed by employing a locally modified version of SCM-AMS program package. \cite{te2001chemistry,ADF2022,ADF2025}
    \item \textit{Analysis and refinement:} 
    For each system, the final QM/FQ$^{(a,b)}$ and QM/FQF$\mu$ excitation energies are obtained by averaging the values computed for the 200 snapshots.  
\end{enumerate}

\section{Results and Discussion}

In this section, vertical excitation energies of ACRO, ACE, CAFF, DOXO, BET, PNA, and C153 (see \cref{fig:structures}) in aqueous solution are discussed. 
Such molecules exhibit different types of electronic transitions.  ACRO and ACE are selected as representative systems for $n\rightarrow \pi^*$ transitions, due to their pronounced solvatochromic shifts and extensive previous studies. \cite{skoko2023towards,aidas2011fluorescence,catalan2011solvatochromism,renge2009solvent} CAFF and DOXO are chosen for $\pi \rightarrow \pi^*$ transitions, whereas BET, PNA and C153 are selected to assess the performance of the methods in describing CT transitions. Most of these molecules have been previously investigated under both LR and SS solvent regimes, providing a solid basis for comparison. \cite{marenich2011practical,marenich2015electronic,giovannini2019electronic,loco2016qm,ambrosetti2021quantum,olszowka2016computational,egidi2018nature,giovannini2019simulating}

Computed values are obtained by coupling QM/FQF$\mu$ with different solvent regimes, namely GSRF, LR, cLR, cLR$^2$, and VEM. The corresponding QM/FQ values, as obtained employing two different parametrizations \cite{rick1994dynamical,giovannini2019effective} are given in Sec. S4 in the SM. For each molecule, 200 snapshots extracted from MD trajectories are considered, each yielding different signals. These variations reflect the flexibility of the solute and the differing solvent distributions around it, reproducing the solvent-induced band inhomogeneous broadening. Additionally, 200 snapshots are sufficient to achieve convergence, as shown in Tab. S1 in the SM.

\subsection{Analysis of electronic transitions}

Before focusing on excitation energy values, the nature of the involved electronic transitions is analyzed, and GS, ES, and transition dipole moments are discussed to highlight the potential CT character of these transitions. 

\begin{table}[!htbp]
\caption{QM/FQF$\mu$ ground state ($\mu_{GS}$), excited state ($\mu_{ES}$) and transition ($\mu^T$) dipole moments (Debye). $\mu_{ES}$ are calculated using the relaxed density matrix obtained via the LR approach. All values represent averages over 200 snapshots.}
\centering
\label{t:dipole_fqfmu}
\begin{tabular}{ |c|c|c|c| } 
 \hline
 Molecules & $\mu_{GS}$ (D) & $\mu_{ES}$ (D) & $\mu^{T}$ (D)\rule{0pt}{2.5ex} \\ 
 \hline
 ACRO & 5.9 & 3.1 & 0.2 \\ 
 ACE & 5.6 & 4.1 & 0.1 \\ 
 CAFF & 6.2 & 5.7 & 3.6 \\ 
 DOXO & 220.1 & 220.1 & 5.9 \\
 BET & 30.2 & 14.5 & 3.0 \\ 
 PNA & 12.3 & 19.1 & 5.4 \\ 
 C153 & 14.3 & 22.7 & 5.9 \\ 
 \hline
\end{tabular}
\end{table}

\begin{figure*}[!t]
    \centering
    \includegraphics[width=\textwidth]{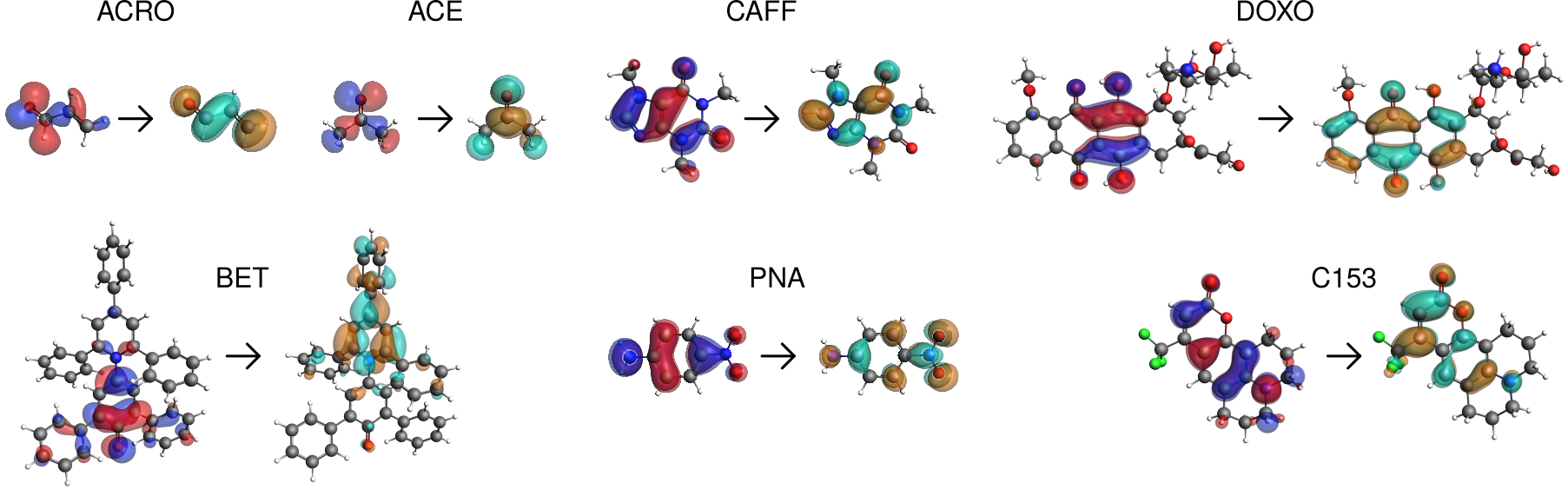}
    \caption{HOMO (red and blue) and LUMO (yellow and green) orbitals involved in the studied electronic transitions of acrolein (ACRO), acetone (ACE), caffeine (CAFF), doxorubicin (DOXO), betaine-30 (BET), para-nitroaniline (PNA), and coumarin 153 (C153).}
    \label{fig:transitions}
\end{figure*}

For ACRO and ACE, the HOMO–LUMO transition corresponds to a $n \rightarrow \pi^*$ excitation. As shown in \cref{fig:transitions}, for ACRO the HOMO is localized on the nonbonding oxygen orbital (\(n_{\mathrm{O}}\)), while the LUMO spans the antibonding $\pi^*$ orbitals of the carbonyl group (\(\pi^*_{\mathrm{C=O}}\)) and the C=C double bond (\(\pi^*_{\mathrm{C=C}}\)). Similarly, in the case of ACE, the HOMO is primarily composed of \(n_{\mathrm{O}}\), and the LUMO includes the antibonding \(\pi^*_{\mathrm{C=O}}\). According to QM/FQ\(\mu\) data reported in \cref{t:dipole_fqfmu}, the dipole moment (computed by using the relaxed density) decreases moving from the GS to the ES, as electrons move from peripheral nonbonding orbitals to central antibonding \(\pi^*\) orbitals. Specifically, ACE exhibits a moderate decrease in dipole moment (5.60 D to 4.1 D, 26\%), while ACRO undergoes a more significant reduction (5.9 D to 3.1 D, 47\%). These results reflect an intramolecular CT from the \(n_{\mathrm{O}}\) orbital to the antibonding \(\pi^*\) orbitals. In ACE, this CT remains confined to the carbonyl group, while in ACRO, it extends to the \(\pi^*\) orbital of the C=C double bond. Furthermore, these \(n \rightarrow \pi^*\) transitions are dark, as shown by their near-zero transition dipole moments (\(\mu^T\)) reported in \cref{t:dipole_fqfmu}. These results for ACRO and ACE are consistent with previous studies in the literature \cite{marenich2011practical,egidi2021polarizable,aidas2011fluorescence,marenich2015electronic,goletto2021combining,giovannini2019quantum}.

Next, we examine CAFF and DOXO, both characterized by a HOMO--LUMO \(\pi \rightarrow \pi^*\) transition. In CAFF the HOMO (see \cref{fig:transitions}) is primarily localized on the \(\pi_{\mathrm{C=C}}\) orbital of the shared carbon atoms between the rings, with a minor contribution from the nitrogen and oxygen lone pairs (\(n_{\mathrm{N}}\), \(n_{\mathrm{O}}\)). The LUMO is dominated by \(\pi^*\) antibonding orbitals spread across the rings and carbonyl groups. \cite{gomez2024caffeine,gomez2020absorption,singh2014spectroscopic} For DOXO, the HOMO--LUMO transition occurs within the anthracycline chromophore (see \cref{fig:transitions}), consistent with previous studies in the literature.\cite{olszowka2016computational} QM/FQF\(\mu\) calculations reported in \cref{t:dipole_fqfmu} show minimal changes in the dipole moment upon going from the GS to the ES, with CAFF decreasing from 6.2 D to 5.7 D, while DOXO remains constant at 220.1 D. These findings are consistent with previous studies,\cite{gomez2020absorption,egidi2018nature} which report small DCT values \cite{le2011qualitative} of approximately 1.8 \AA{} for both molecules, which confirm the limited CT nature of these transitions. The bright \(\pi \rightarrow \pi^*\) transitions in CAFF and DOXO are associated with large transition dipole moments (\(\mu^T\)), reaching 3.6~D for CAFF and 5.9~D for DOXO.

BET, PNA, and C153 are characterized by CT transitions of  \(n \rightarrow \pi^*\) or \(\pi \rightarrow \pi^*\) nature. In BET, as shown in \cref{fig:transitions}, the HOMO is localized on the phenolate moiety, while the LUMO is located on the pyrimidine fragment, resulting in intramolecular CT between the two regions upon excitation, substantially reducing the zwitterionic nature of the GS.\cite{ambrosetti2021quantum,loco2016qm} According to \cref{t:dipole_fqfmu}, BET shows the most significant dipole moment reduction, from 30.2 D in the GS to 14.5 D in the ES (52\%  drop). For PNA, the HOMO is mainly composed of the nitrogen lone-pair \((n_{\mathrm{N}})\) of the amino group, combined with the \(\pi\) orbitals of the benzene ring and the nitro group (see \cref{fig:transitions}). The LUMO, in contrast, predominantly consists of \(\pi^*\) orbitals on the benzene ring and the nitro group. This intramolecular CT shifts the electron density from the donor amino group to the acceptor nitro group, increasing the dipole moment from 12.3 D in the GS to 19.1 D in the ES (+55\%).
In C153, the HOMO is delocalized across the entire molecule, with significant contributions from the ``central'' benzene ring and the nitrogen atom, as depicted in \cref{fig:transitions}. The LUMO, on the other hand, is primarily localized on the ``quinone-like'' terminal ring, with a strong contribution from the carbonyl group. \cite{improta2006state} As a result, the dipole moment increases even more substantially than in PNA, from 14.3 D in the GS to 22.7 D in the ES (58\% increase). 
All three systems show bright CT transitions, as evidenced by their large transition dipole moments $\mu^T$: 3.0 D for BET, 5.4 D for PNA, and 5.9 D for C153.

All transitions discussed above, and corresponding dipole moments, are obtained using the LR approach. In Tab. S4 in the SM, values calculated with the relaxed density matrix obtained by the VEM approach are reported. Overall, the differences between LR, cLR, and VEM are minimal, aligning with the observations and trends previously discussed.

The role of including dipoles as polarization variables in the polarizable classical portion can be evaluated by comparing the previous results with the corresponding values computed with QM/FQ, employing two parameterizations sets, namely FQ$^{a}$\cite{rick1994dynamical} and FQ$^{b}$\cite{ambrosetti2021quantum} (see Tabs. S2-S3, S5-S6 in the SM). In general, QM/FQ calculations yield dipole moments that are slightly lower than the corresponding QM/FQF$\mu$ values, with only a few exceptions. Notably, the choice of solvent model and response regime has a negligible impact on the computed dipole moments, and the nature of the transition remains consistent across all approaches.

\subsection{Excitation energies and solvatochromic shifts}

\begin{figure}[h!]
    \centering
            \includegraphics[width=0.45\textwidth]{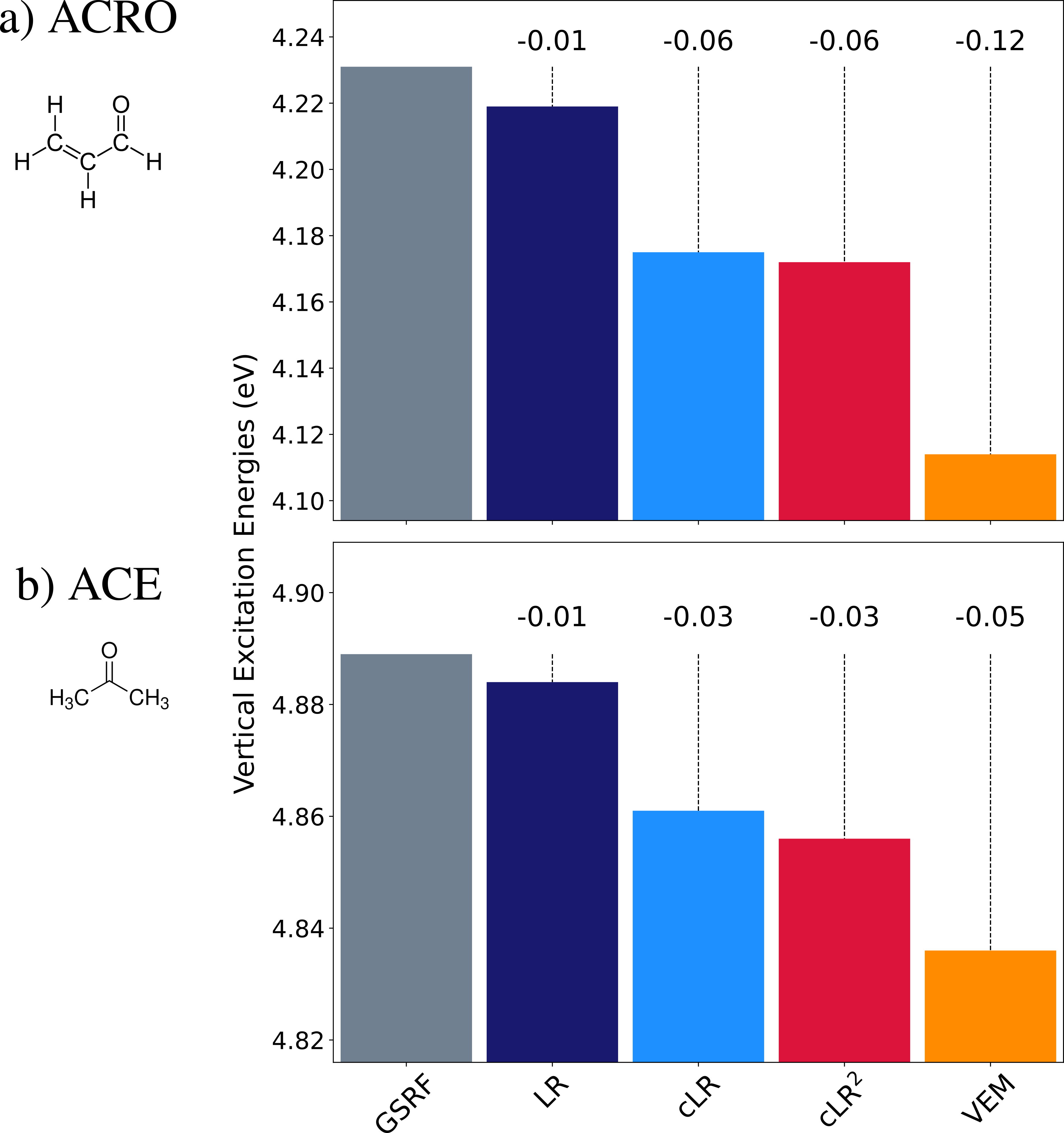}
        \caption{QM/FQF$\mu$ vertical excitation energies of (a) ACRO and (b) ACE, averaged over 200 snapshots. Data refer to the HOMO–LUMO $n \to \pi^*$ transition, under different solvent regimes (GSRF, LR, cLR, cLR$^2$, and VEM). The dashed line indicates the correction to the GSRF vertical excitation energy.}
        \label{fig:acro_ace}
\end{figure}

Moving to QM/FQF$\mu$ vertical excitation energies, for ACRO and ACE, characterized by a $n \rightarrow \pi^*$ transition, $\omega_{\text{GSRF}}$ and $\omega_{\text{LR}}$ are remarkably similar, as illustrated in \cref{fig:acro_ace}. In particular, QM/FQF$\mu$ values for $\omega_{\text{GSRF}}$ and $\omega_{\text{LR}}$ are 4.23 eV and 4.22 eV for ACRO, and 4.89 eV and 4.88 eV for ACE, respectively, demonstrating an effect resulting from the LR regime of maximum -0.01 eV. This small deviation can be attributed to the low transition density associated with the HOMO--LUMO $n \rightarrow \pi^*$ transitions
, which results in a small dynamic response of the solvent. \cite{caricato2006formation,caricato2014corrected} The QM/FQF$\mu$ SS corrections to $\omega_{\text{GSRF}}$ introduced via the cLR and cLR$^2$ approaches (i.e., $\omega_{\text{cLR}} - \omega_{\text{GSRF}}$ and $\omega_{\text{cLR$^2$}} - \omega_{\text{GSRF}}$) are -0.06 eV and -0.03 eV for ACRO and ACE, respectively, as reported in \cref{fig:acro_ace}. These corrections are significantly larger than those obtained with the LR approach, particularly for ACRO, which aligns with our observations of substantial dipole moment changes from the ground to the excited state and is consistent with findings reported in previous studies. \cite{caricato2006formation,loco2016qm,giovannini2019quantum,giovannini2019electronic} 
The $\text{cLR}^2$ approach accounts for both SS and LR effects and yields results similar to those of the cLR method due to the minimal impact of the LR contribution.
When employing a fully SS method such as VEM, the corrections to \(\omega_{\text{GSRF}}\) (i.e., \(\omega_{\text{VEM}} - \omega_{\text{GSRF}}\)) are -0.12 eV for ACRO and -0.05 eV for ACE. Thus, for transitions of this nature, the solvent response associated with ES relaxation significantly exceeds its dynamical counterpart. Furthermore, the disparity between cLR (and cLR$^2$) and VEM corrections highlights the limitations of the first-order perturbative contribution of cLR, particularly for molecules with significant electronic density rearrangements. While cLR or cLR$^2$ approaches may be sufficient for systems with a modest difference between GS and ES electronic densities, the inclusion of VEM ensures a more accurate and comprehensive recovery of the SS solvent response.
This trend is consistent with observations reported in the case of VEM/PCM calculations \cite{marenich2011practical}. 

\begin{figure}[h!]
    \centering
            \includegraphics[width=0.45\textwidth]{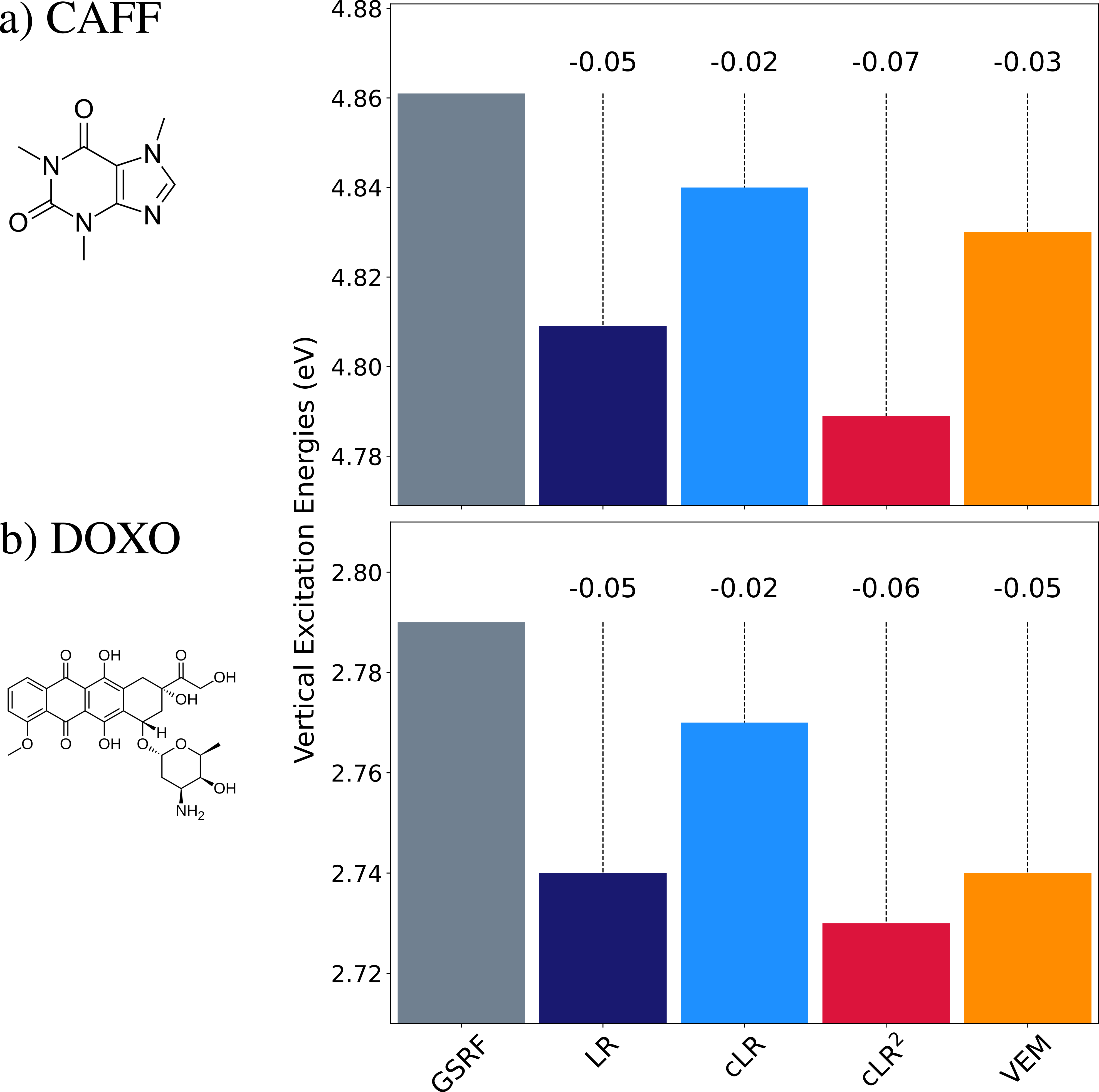}
        \caption{QM/FQF$\mu$ vertical excitation energies of (a) CAFF and (b) DOXO, averaged over 200 snapshots. Data refer to the HOMO–LUMO $\pi \to \pi^*$ transition, under different solvent regimes (GSRF, LR, cLR, cLR$^2$, and VEM). The dashed line indicates the correction to the GSRF vertical excitation energy.}
        \label{fig:caff_doxo}
\end{figure}

Bright $\pi \rightarrow \pi^*$ HOMO–LUMO transitions characterize CAFF and DOXO. For CAFF, $\omega_{\text{GSRF}}$ and $\omega_{\text{LR}}$ excitation energies are 4.86 eV and 4.81 eV, respectively, while the corresponding values for DOXO are 2.79 eV and 2.74 eV, with a LR correction of -0.05 eV, reflecting the solvent response to the transition density (see \cref{fig:caff_doxo}).
QM/FQF$\mu$ SS corrections to $\omega_{\text{GSRF}}$ obtained using the cLR approach (i.e., $\omega_{\text{cLR}} - \omega_{\text{GSRF}}$) are particularly small: $-0.02$ eV for both CAFF and DOXO, as shown in \cref{fig:caff_doxo}. The cLR$^2$  method, which incorporates both LR and SS effects, yields excitation energies of 4.79 eV for CAFF and 2.73 eV for DOXO, with corrections of -0.07 eV and -0.06 eV, respectively, similar to the LR case. The larger magnitude of LR corrections compared to cLR is expected, given that $(\mu_{ES} - \mu_{GS})^2 < 2 (\mu^{T})^2$. \cite{caricato2006formation} VEM excitation energies deviate by -0.03 eV for CAFF and -0.05 eV for DOXO from $\omega_{\text{GSRF}}$. Together, these results emphasize the dominant role of the solvent dynamic response in determining excitation energies, with SS effects remaining relatively minor.
  
\begin{figure}[h!]
    \centering
            \includegraphics[width=0.45\textwidth]{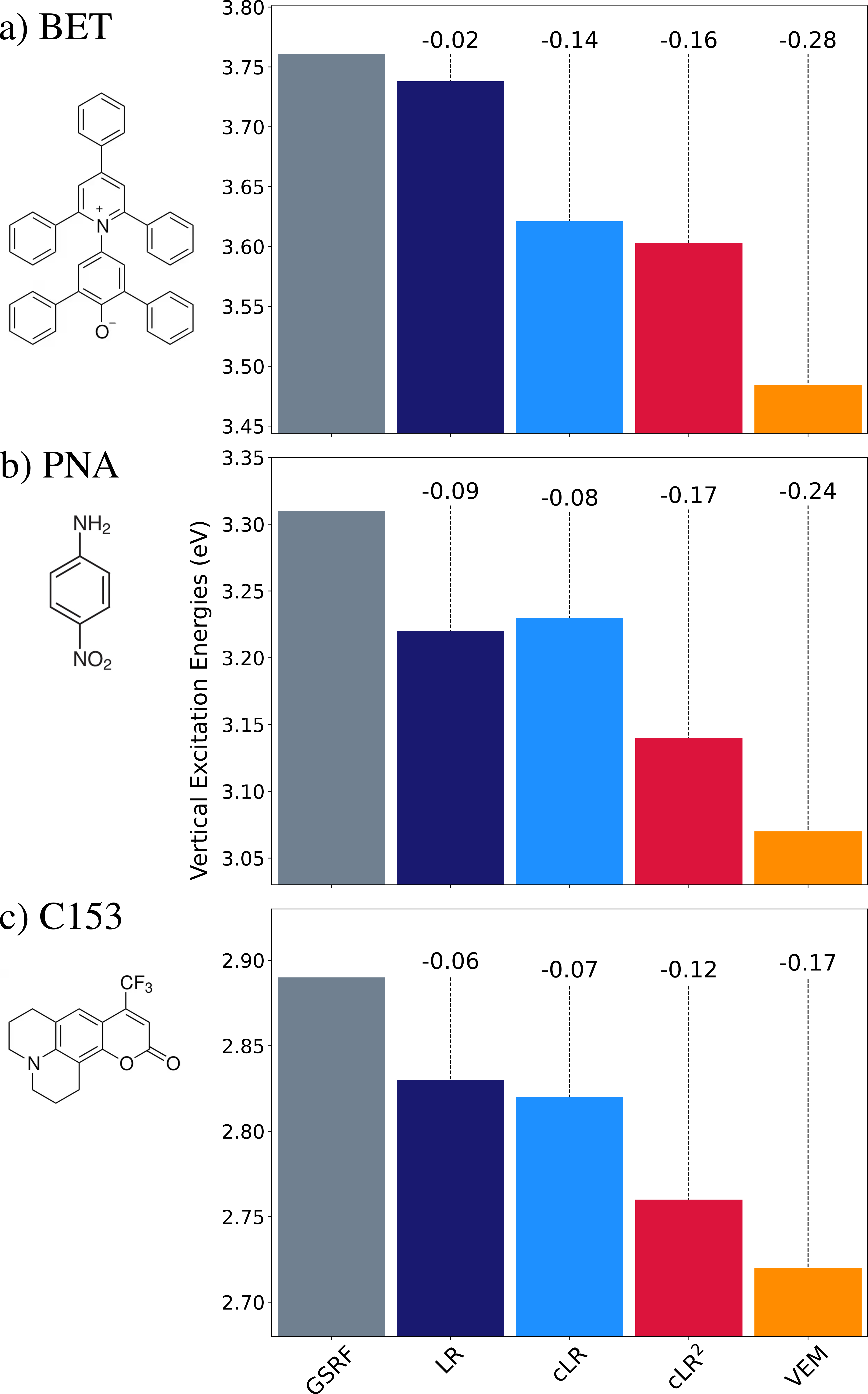}
        \caption{QM/FQF$\mu$ vertical excitation energies of (a) BET, (b) PNA, and (c) C153 averaged over 200 snapshots for the HOMO–LUMO CT transitions. Data refer to different solvent regimes (GSRF, LR, cLR, cLR$^2$, and VEM). The dashed line indicates the correction to the GSRF vertical excitation energy.}
        \label{fig:bet_pna_c153}
\end{figure}

BET, PNA, and C153 are characterized by large transition dipole moments and a significant change in the electronic density upon CT excitation. The computed excitation energies are reported in \cref{fig:bet_pna_c153}.
In the case of BET, $\omega_{\text{GSRF}}$ and $\omega_{\text{LR}}$ are similar, with QM/FQF$\mu$ values of 3.76 eV and 3.74 eV, respectively, showing a LR correction of -0.02 eV.  $\omega_{\text{cLR}} - \omega_{\text{GSRF}}$ is -0.14 eV, while $\omega_{\text{cLR}^2} - \omega_{\text{GSRF}}$ gives -0.16 eV, thus highlighting a significant SS effect. The ($\omega_{\text{VEM}} - \omega_{\text{GSRF}}$) correction substantially increases to -0.28 eV.
The pronounced SS solvent effect observed for BET is primarily driven by the CT nature of the transition between the two molecular fragments, which causes a significant redistribution of the electronic density upon excitation. 
For PNA, the $\omega_{\text{GSRF}}$ and $\omega_{\text{LR}}$ excitation energies are 3.31 eV and 3.22 eV, respectively, with an LR correction of -0.09 eV. For C153, the $\omega_{\text{GSRF}}$ is 2.89 eV and $\omega_{\text{LR}}$ is 2.83 eV, corresponding to a correction of -0.06 eV. The $\omega_{\text{cLR}}$ energies are 3.23 eV for PNA and 2.82 eV for C153, with corrections relative to $\omega_{\text{GSRF}}$ ($\omega_{\text{cLR}} - \omega_{\text{GSRF}}$) of $-0.08$ eV and $-0.07$ eV, respectively. For PNA, the larger LR correction compared to the cLR correction is consistent with expectations, as $(\mu_{ES} - \mu_{GS})^2 < 2 (\mu^{T})^2$. \cite{caricato2006formation} In contrast, the opposite trend is observed for C153.
The LR effect is significant because the large transition dipole moments of these systems enhance the dynamic solvent response. Simultaneously, the SS effect plays an important role due to the CT nature of the transitions. These effects are combined in the $\omega_{\text{cLR}^2}$ model, yielding excitation energies of 3.14 eV for PNA and 2.76 eV for C153, with corrections to $\omega_{\text{GSRF}}$ of $-0.17$ eV and $-0.12$ eV, respectively.
By capturing the full SS effect with the VEM approach, the corrections to $\omega_{\text{GSRF}}$ ($\omega_{\text{VEM}} - \omega_{\text{GSRF}}$) amount to $-0.24$ eV for PNA and $-0.17$ eV for C153. Thus, the corrections introduced by VEM for BET, PNA, and C153 are particularly large. Notably, for C153 and PNA, both the solvent’s response to the ES relaxation and the transition density are particularly relevant. Remarkably, the first-order perturbative correction provided by the cLR approach does not fully capture the SS contribution, which is instead accurately recovered by the VEM approach.

The role of including dipoles as polarization variables in the polarizable classical portion can also be evaluated in this case by comparing the previous results with the corresponding values computed with QM/FQ, employing the two parameterization sets FQ$^{a}$\cite{rick1994dynamical} and FQ$^{b}$\cite{ambrosetti2021quantum}.
Values are given in Figs. S2-S7 in the SM. QM/FQ$^{a}$ and QM/FQ$^{b}$ vertical excitation energies are lower than QM/$\text{FQF}\mu$ values, and in particular $\omega_{\text{QM/FQ}^a} < \omega_{\text{QM/FQ}^b} < \omega_{\text{QM/FQF}\mu}$ for transitions $n \to \pi^*$ transitions (ACRO, ACE, and BET). The opposite trend is instead observed for $\pi \to \pi^*$ transitions (CAFF, DOXO, PNA, and C153). These trends highlight that including polarizable dipoles to describe solute-solvent polarizations enhances solvent effects, similarly to previously studied cases.\cite{gomez2024modeling,nicoli2022assessing} Furthermore, the overall trends can be rationalized by considering the physical description provided by each model: QM/FQF$\mu$\cite{giovannini2019polarizable} and QM/FQ$^{b}$\cite{ambrosetti2021quantum} are parametrized to correctly reproduce high-level electrostatic (and polarization) solute-solvent interactions, while QM/FQ$^{a}$ is specifically designed for bulk water.\cite{rick1994dynamical}

\begin{figure*}
    \centering
            \includegraphics[width=0.8\textwidth]{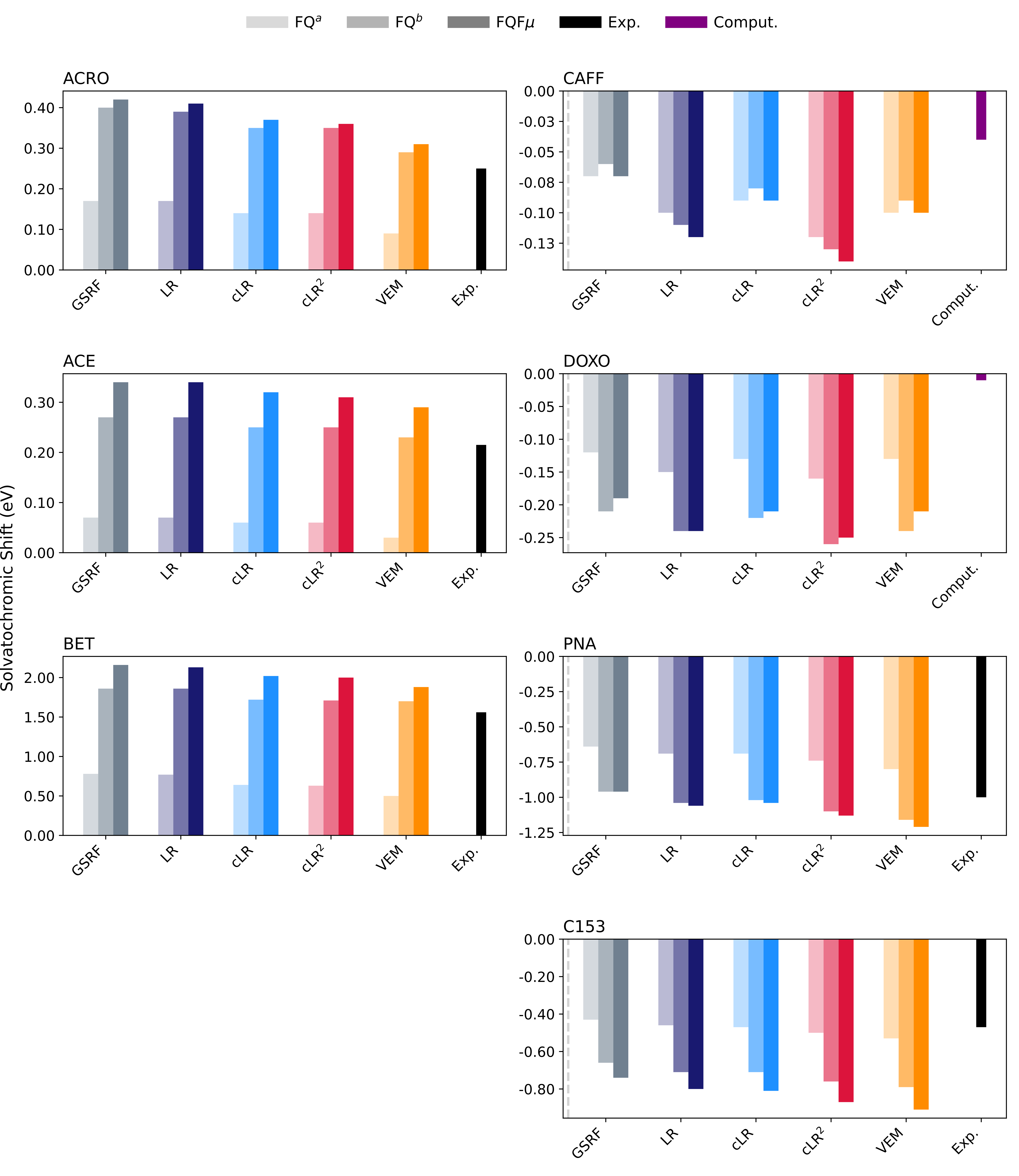}
        \caption{QM/FQ$^{a,b}$ and QM/FQF$\mu$ solvatochromic shifts ($\omega_{model} - \omega_{vac}$) for the investigated molecules (ACRO, CAFF, ACE, DOXO, BET, PNA, C153) computed using different solvent models (GSRF, LR, cLR, cLR$^2$, VEM). 
        Experimental solvatochromic shifts in water are taken from Refs. \citenum{hayes1965solvent,bayliss1954solvent,bayliss1968solvent,catalan2011solvatochromism,renge2009solvent} for ACE, \citenum{moskvin1966experimental} for ACRO, \citenum{reichardt2011solvents} for BET, \citenum{kovalenko2000femtosecond} for PNA, and \citenum{horng1995subpicosecond} for C153 (in DMSO). Computational data are taken from Ref. \citenum{rijal2022quantum} for CAFF and Ref. \citenum{nicoli2022assessing} for DOXO.}
        \label{fig:shift}
\end{figure*}

The best way to compare computed and experimental transition energies is by focusing on gas-to-solution solvatochromic shifts. This way, the systematic error connected to a specific choice of the QM level (DFT functional and basis set) is reduced, and the quality of the description of solvent effects is highlighted.\cite{giovannini2019quantum,loco2016qm,nicoli2022assessing} \cref{fig:shift} and Tabs. S10-S12 in the SM show QM/FQ$^{a,b}$ and QM/FQF$\mu$ solvatochromic shifts ($\omega_{\text{QM/FQ(F}\mu)} - \omega_{\text{vac}}$), obtained with the diverse solvent regimes.

For  $n \to \pi^*$ transitions (ACRO and ACE), experimental vacuum-to-water solvatochromic shifts are positive, indicating a blueshift, with reported values of 0.25 eV \cite{moskvin1966experimental} and 0.19–0.22 eV \cite{hayes1965solvent,bayliss1954solvent,bayliss1968solvent,catalan2011solvatochromism,renge2009solvent}, respectively. QM/FQ$^a$ calculations substantially underestimate these shifts (see \cref{fig:shift}), and no improvement follows from the inclusion of SS corrections. In contrast, QM/FQ$^b$ and QM/FQF$\mu$ tend to overestimate the experimental shifts. However, the use of SS models such as VEM reduces the discrepancy, yielding solvatochromic shifts of 0.29 eV (QM/FQ$^b$) and 0.31 eV (QM/FQF$\mu$) for ACRO, and 0.23 eV (QM/FQ$^b$) and 0.29 eV (QM/FQF$\mu$) for ACE, respectively. Among the studied models, QM/FQ$^b$ provides the best agreement with experiment.

The experimental solvatochromic shifts are not reported in the literature for the $\pi \to \pi^*$ transitions of CAFF and DOXO. For CAFF, a theoretical study reported a shift of $-0.04$ eV \cite{rijal2022quantum}, while for DOXO, we extrapolated a solvatochromic shift of $-0.01$ eV in a previous work.\cite{nicoli2022assessing} Our calculations overestimate both values, consistent with our previous findings at the LR level.\cite{nicoli2022assessing}

Moving to CT transitions, a notably large positive experimental solvatochromic shift of 1.56 eV was reported for BET\cite{reichardt2011solvents}.
As in the case of ACRO and ACE, QM/FQ$^a$ shifts underestimate the experimental excitation energies, and the inclusion of SS corrections does not lead to any improvement over the LR results. In contrast, LR calculations based on QM/FQ$^b$ and QM/FQF$\mu$ overestimate the excitation energies (1.86 eV and 2.13 eV, respectively). However, the agreement with experimental values improves significantly when SS effects are accounted for—particularly with the VEM model, resulting in excellent agreement for QM/FQ$^b$ (1.56 eV). 

PNA shows a negative experimental solvatochromic shift of $-1.00$ eV \cite{kovalenko2000femtosecond}. The calculated LR, cLR, and VEM shifts are smaller in absolute value for QM/FQ$^a$, and larger for QM/FQ$^b$ and QM/FQF$\mu$. Overall, all computed values show reasonable agreement with experiment, with VEM yielding the most accurate result for QM/FQ$^a$, and cLR providing the closest match for both QM/FQ$^b$ and QM/FQF$\mu$. In contrast, C153 shows an experimental negative shift of $-0.47$ eV in DMSO \cite{horng1995subpicosecond}. Since water is more polar than DMSO, the experimental vacuum-to-water shift - although undocumented in the literature - is expected to be even more negative. Therefore, the calculated shifts overestimate the experimental value in DMSO, with smaller deviations observed for QM/FQ$^a$ and larger ones for QM/FQ$^b$ and QM/FQF$\mu$.

The general overestimation of solvatochromic shifts by QM/FQ$^b$ and QM/FQF$\mu$ explicit solvent models is not unexpected \cite{nicoli2022assessing,loco2016qm,giovannini2019quantum} and can be due to the lack of inclusion of solute-solvent non-electrostatic interactions, which balance electrostatic contributions and bring the computed values closer to the experimental ones.\cite{giovannini2019quantum,gomez2024close,ambrosetti2021quantum} Also, computed shifts refer to vertical excitation energies, whereas experimental values are extracted from spectral absorption maxima. The discrepancy between the two quantities may also be due to vibronic progressions. Overall, however, among all tested approaches, QM/FQ$^b$ provides the best agreement with the experimental values.

\section{Conclusion and future perspectives}

In this study, the QM/FQ and QM/FQF$\mu$ approaches have been formulated within the fully self-consistent, state-specific VEM framework to simulate solvent responses to changes in solute density upon electronic excitation. While the LR approach captures the dynamic response of the solvent to the QM transition density, it does not account for solvent relaxation to the solute’s redistributed charge.
To systematically investigate solvent effects across different electronic transitions, vertical excitation energies were computed under various solvent response regimes: GSRF, LR, cLR, cLR$^2$, and VEM. This strategy allows a detailed analysis of solvent response for $n \rightarrow \pi^*$, $\pi \rightarrow \pi^*$, and CT excitations.
The results highlight the strong dependence of solvent response on the nature of the electronic transition. Systems undergoing $\pi \rightarrow \pi^*$ transitions, such as caffeine and doxorubicin, are primarily influenced by dynamic solvent effects and can be reasonably described using the LR approach. In contrast, systems with significant CT character, such as betaine-30, p-nitroaniline, and Coumarin 153, as well as those exhibiting $n \rightarrow \pi^*$ transitions like acetone and acrolein, require advanced state-specific corrections to accurately capture solvent relaxation effects. Among the tested methods, VEM emerges as the most reliable approach for modeling these effects.
Comparison with experimental data shows that the QM/FQ$^a$ model consistently underestimates solvatochromic shifts, likely due to its parameterization optimized for bulk water. In contrast, QM/FQ$^b$ and QM/FQF$\mu$ predict larger absolute solvent responses, displaying similar trends. This similarity stems from their shared parameterization based on electrostatic interaction energies, although QM/FQF$\mu$ additionally includes dipoles as a source of polarization.
The overestimation of solvent effects observed for QM/FQ$^b$ and QM/FQF$\mu$ highlights the importance of including non-electrostatic interactions, which play a significant role in modulating the electrostatic solvent response. Developing QM/MM frameworks that consistently incorporate these contributions remains a challenge due to their inherently quantum nature. Future work will focus on extending the models to account for such effects, potentially through specialized QM/MM schemes. \cite{giovannini2017general,giovannini2019effective,giovannini2019quantum,marrazzini2020calculation}
Moreover, the combination of proper solvent response regimes, such as cLR and VEM, with quantum embedding strategies offers a promising direction for further improvement. These strategies could be implemented both within DFT-based approaches \cite{marrazzini2021multilevel,giovannini2023integrated,egidi2021polarizable,wesolowski1993frozen} and high-level correlated methods, such as Coupled Cluster or Complete Active Space, \cite{goletto2021combining,goletto2022linear,sepali2024fully} enabling more accurate modeling of solute–solvent interactions and providing a pathway toward quantitatively reliable predictions of excitation energies in complex environments.

A further future development will involve the formulation of state-specific nuclear gradients within the QM/FQF$\mu$ VEM framework. This advancement will open the way to studying the geometry of molecules in excited states with full state-specific methods, as well as to performing excited-state molecular dynamics simulations.

\begin{acknowledgments}
The authors thank Prof. Ciro Achille Guido (University of Piemonte Orientale, Alessandria, Italy) for useful discussions. CC acknowledge funding from MUR-FARE Ricerca in Italia: Framework per l'attrazione ed il rafforzamento delle eccellenze per la Ricerca in Italia - III edizione. Prot. R20YTA2BKZ and from the European Union's Horizon Europe research and innovation programme under the project HORIZON-MSCA-2023-DN-01 - LUMIÈRE G.A . No 101169312. The Center for High-Performance Computing (CHPC) at SNS is gratefully acknowledged for providing the computational infrastructure.
\end{acknowledgments}

\section*{Data Availability Statement}

The data that support the findings of this study are available from the corresponding author upon reasonable request.

\section*{References}
\bibliography{biblio}

\end{document}